\documentclass[aps,prl,twocolumn,preprintnumbers,superscriptaddress,amsmath,amssymb]{revtex4-1}
\usepackage{graphicx}
\usepackage{subfigure}
\usepackage{epsfig}
\usepackage{dcolumn}
\usepackage{bm}
\usepackage{ulem}
\usepackage{color}

\def\be{\begin{equation}}       \def\ee{\end{equation}}
\def\bea{\begin{eqnarray}}      \def\eea{\end{eqnarray}}
\def\ba{\begin{array} }
\def\ea{\end{array} }
\def\bnum{\begin{enumerate} }
\def\enum{\end{enumerate}}

\def\=>{\Rightarrow}
\def\>{\rightarrow}
\def\A{\uparrow}
\def\V{\downarrow}

\def\eye2{Fathbb{I}}

\renewcommand{\>}{\rangle}

\begin{document}

\title{Possible Triplet $p+ip$ Superconductivity in Graphene at Low Filling}
\author{Tianxing Ma}
\affiliation{Department of Physics, Beijing Normal University, Beijing 100875, China\\}
\affiliation{Beijing Computational Science Research Center, Beijing 100084, China}
\author{Fan Yang}
\thanks{\texttt{yangfan\_blg@bit.edu.cn}}
\affiliation{School of Physics, Beijing Institute of Technology, Beijing, 100081, China}
\affiliation{Beijing Computational Science Research Center, Beijing 100084, China}
\author {Hong Yao}
\thanks{\texttt{yaohong@tsinghua.edu.cn}}
\affiliation{Institute of Advanced Study, Tsinghua University, 100081, China}
\affiliation{Collaborative Innovation Center of Quantum Matter, Beijing, China}
\author {Hai-Qing Lin}
\affiliation{Beijing Computational Science Research Center, Beijing 100084, China}

\begin{abstract}
We study the Hubbard model on the honeycomb lattice with  nearest-neighbor hopping ($t>0$) and next-nearest-neighbor one ($t'<0$). When $t'<-t/6$, the single-particle spectrum is featured by the continuously distributed Van-Hove saddle points at the band bottom, where the density of states diverges in power-law. We investigate possible unconventional superconductivity in such system with Fermi level close to the band bottom by employing both random phase approximation and determinant quantum Monte-Carlo approaches. Our study reveals a possible triplet $p+ip$  superconductivity in this system with appropriate interactions. 
 Our results might provide a possible route to look for triplet superconductivity with relatively-high transition temperature in a low-filled graphene and other similar systems.
\end{abstract}

\maketitle

\section{introduction}
 Graphene, a single layer of carbon atoms forming a honeycomb lattice, has been among the most exciting research fields since sythesized\cite{graphene}. Enormous attentions on this remarkable material have been focused on exploring physics related to its Dirac-cone band structure\cite{dirac}. For graphene close to half-filling, the density of states (DOS) at the Fermi level is almost vanishing; as a consequence, relatively weak/intermediate short-range repulsive interactions in general do not induce phase transitions at low temperature\cite{dirac}.  Nonetheless, exotic phases might be induced by repulsive interactions when the Fermi level is finitely away from the Dirac point. For instance, it was shown by renormalization group (RG) calculations that unconventional/topological superconductivity (SC) is induced by weak repulsive interactions in honeycomb Hubbard models finitely away from half-filling\cite{raghu10, Nandkishore}. More recently, exotic phases such as $d+id$ \cite{raghu10,chubukov,wang,thomale,did1,did2,did3,Platt} topological superconductivity\cite{Qi-Zhang,Hasan-Kane} and Chern band insulators with spin density waves\cite{litao11,martin08} near the type-I Van-Hove singularity (VHS) at 1/4 electron or hole doping, where the DOS at Fermi level diverges logarithmically. Such logarithmically diverging DOS close the VHS may significantly raise superconducting transition temperature. More recently, it was shown by RG analysis that topological triplet $p+ip$ superconductivity can generically occur in systems at type-II VHS where the saddle points are not at time-reversal-invariant momenta\cite{yao13,chenxi}.


In 2D, for a Fermi surface with discrete Van-Hove saddle points, the DOS at Fermi level diverges only logarithmically. It would be interesting to study phases in systems with a power-law diverging DOS. Indeed, it was shown that for the hopping parameters satisfying $t'<-t/6$, an inverse-square-root diverging DOS occurs close to band bottom of the lower band, where the band bottom is a closed line instead of discrete points as shown in Fig. \ref{Fig:Sketch} (b). In the graphene, such hopping parameters are possible\cite{FP,exp1,exp2}, and high levels of doping are experimentally accessible recently\cite{doping_high}. Note that the band bottom occurring at a closed line only when no third-neighbor or longer-range hopping is considered. This kind of line band bottom may be considered as a set of continuously distributed VH saddle points.
Recent determinant quantum Monte-Carlo (DQMC) study has revealed ferromagnetic-like spin-correlations in such system\cite{Ma2010}, which implies possibility of a dominant triplet pairing state in this system with repulsive interactions. 

In this paper, we report both random phase approximation (RPA) analysis and DQMC studies of pairing symmetries of possible SC induced by weak or intermediate repulsive interactions in graphene at low fillings whose DOS at Fermi level is significantly enhanced by the power-law singularity at the band bottom. 
Both numerical approaches obtain the $p+ip$ triplet pairing as the leading instability of the system in different parameter regimes. For $t'=-0.2t$, $U/t=3.0$, and filling $n=0.2$, the transition temperature $T_{c,\textrm{triplet}}$ into the triplet pairing state is estimated to be in the order of $10^{-2}t$. For graphene $t\sim 2.0$eV, this implies that the $T_{c,\textrm{triplet}}$ in graphene might be as high as 200K when the Fermi level is tuned appropriately close to the band bottom.
These results might provide a possible route to look for triplet superconductivity with relatively-high transition temperature in graphene at low filling.

\section{Model and approach}
 We start from the following Hubbard model on the honeycomb lattice
\begin{equation}
H=-t\sum_{\left\langle i,j\right\rangle}c^{\dagger}_{i\sigma}c_{j\sigma}-
t'\sum_{\left\langle\langle i,j\right\rangle\rangle}c^{\dagger}_{i\sigma}c_{j\sigma}+U\sum_{i}n_{i\uparrow}n_{i\downarrow}, \label{model}
\end{equation}
where $c^\dag_{i\sigma}$ is the electron creation operator at site $i$ and with spin polarization $\sigma=\A,\V$ and $U$ labels the on-site repulsive interaction. Here the $t$ and $t'$ terms describe the nearest neighbor (NN) and next nearest neighbor (NNN) hoppings, respectively. We consider the case of $t>0$ and $t'<0$, which is supported by recent first principle calculations\cite{FP} and experiments\cite{exp1}. As the ratio $|\frac{t'}{t}|$ varies from around 0.1\cite{exp1} to around 0.3\cite{exp2} in different experiments, we focus on the possible cases with $t'<-t/6$ and take $t'=-0.2t$ in our calculations unless stated otherwise.

\begin{figure}[tbp]
\includegraphics[scale=0.4]{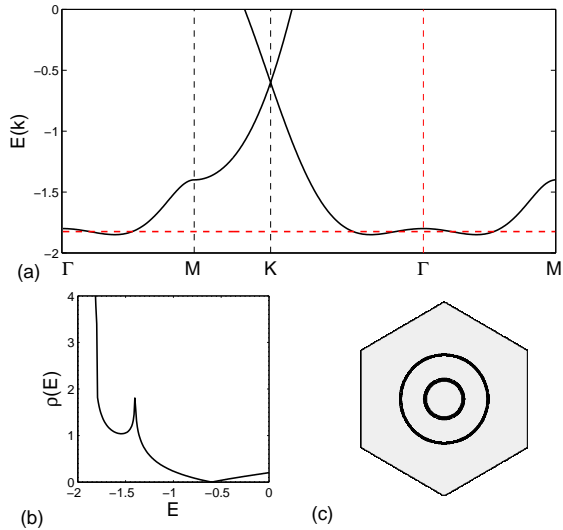}
\caption{(Color online)(a) The energy band along high symmetry line in the first Brillouin Zone; (b) The DOS as function of energy with $t'=-0.2t$; and (c) The Fermi surface at filling $n=0.2$.}
\label{Fig:Sketch}
\end{figure}

The  band structure is shown in Fig. \ref{Fig:Sketch}(a), together with the Fermi levels for filling $n=0.2$ per site. We notice one remarkable feature of this band structure: the band bottom of this system does not locate at the $\Gamma$-point; instead it consists two closed lines around $\Gamma$. As a consequence, the DOS is divergent in an inverse-square-root fashion near the band bottom, as shown in Fig. \ref{Fig:Sketch}(b). The Fermi surface (FS) of the system at $n=0.2$ is shown in Fig. \ref{Fig:Sketch}(c), which contains an inner hole-pocket and an outer electron-pocket. Such a Hubbard-model with only on-site interaction has been widely engaged\cite{chubukov,wang,thomale,did2,did3} to describe the graphene doped to near the VH points because at such dopings, the divergent DOS on the FS leads to strong screening of the Coulomb interaction.

In the following, we adopt perturbative RPA analysis for weak $U$ interactions and the DQMC calculations for relatively strong $U$ to investigate the pairing symmetries of the possible SC at low filling.

\section{RPA treatment}
We adopted the standard multi-orbital RPA approach\cite{Scalapino89,Scalapino-RMP,RPA1,Kubo2007,Graser2009,RPA2} in our study for the small $U$ ($=0.1t$) case.

\begin{figure}[tbp]
\includegraphics[scale=0.4]{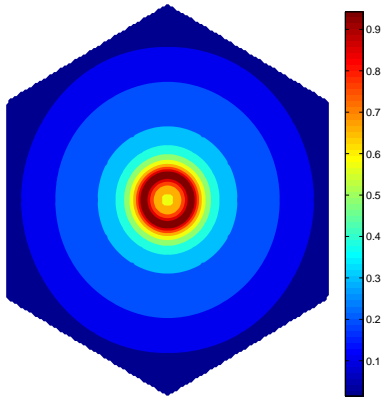}
\caption{(Color online) Largest eigenvalues of the susceptibility matrix in non-interacting limit in the first Brillouin-Zone.}
\label{Fig:Spin}
\end{figure}

Various susceptibilities of non-interacting electrons of this system are defined as
\begin{eqnarray}
 \chi^{(0)l_{1},l_{2}}_{l_{3},l_{4}}\left(\mathbf{q},\tau\right)\equiv
 \frac{1}{N}\sum_{\mathbf{k_{1},k_{2}}}\left<T_{\tau}c^{\dagger}_{l_{1}}(\mathbf{k_{1}},\tau)
 c_{l_{2}}(\mathbf{k_{1}+q},\tau)\right.\nonumber\\
 \left.c^{+}_{l_{3}}(\mathbf{k_{2}+q},0)c_{l_{4}}(\mathbf{k_{2}},0)\right>_0,\label{free_sus}
 \end{eqnarray}
 where $l_{i}$ $(i=1,2)$ denotes orbital (sublattice) index. 
Largest eigenvalues of the susceptibility matrix $\chi^{(0)}_{l,m}\left(\mathbf{q}\right)\equiv
\chi^{(0)l,l}_{m,m}\left(\mathbf{q},i\nu=0\right)$ is shown in
Fig. \ref{Fig:Spin} for filling $n=0.1$, which shows dominant distributions on a small circle around the $\Gamma$-point. This suggests strong ferromagnetic-like intra-sublattice spin fluctuations in the system. Generally, it is found that at low fillings, the radius of the circle scales with filling. At low fillings, the eigenvector of the susceptibility matrix reveals that the inter-sublattice spin fluctuations in the system are also ferromagnetic-like, although somewhat weaker than the intra-sublattice ones. Such ferromagnetic-like spin fluctuations are consistent with the ferromagnetic spin correlations revealed by the DQMC calculations\cite{Ma2010}.

With weak Hubbard-$U$, the spin ($\chi^{s}$)
or charge ($\chi^{c}$) susceptibilities in the RPA level are given by
\begin{equation}
\chi^{s\left(c\right)}\left(\mathbf{q},i\nu\right)=
\left[I\mp\chi^{(0)}\left(\mathbf{q},i\nu\right)\bar U\right]^{-1}\chi^{(0)}
\left(\mathbf{q},i\nu\right),\label{RPA}
\end{equation}
where $\bar U^{\mu\nu}_{\mu'\nu'}$ ($\mu\nu=1,2$) is a 4$\times$4 matrix, whose only two nonzero
elements are $\bar U^{11}_{11}=\bar U^{22}_{22}=U$. Clear, the repulsive Hubbard-$U$ suppresses $\chi^{c}$ but enhances $\chi^{s}$. Thus, the spin fluctuations take the main role of mediating the cooper pairing in the interacting system\cite{Scalapino89}. In the RPA level, the cooper pairs near the FS acquire an effective interaction $V_\textrm{eff}$\cite{Scalapino89,Scalapino-RMP,RPA2} via exchanging the spin fluctuations represented by the spin susceptibilities. From this effective interaction, one obtains the linearized gap equation near the superconducting critical temperature $T_c$, solving which one obtains the leading pairing symmetry (symmetries) of the system.

\begin{figure}[tbp]
\includegraphics[scale=0.45]{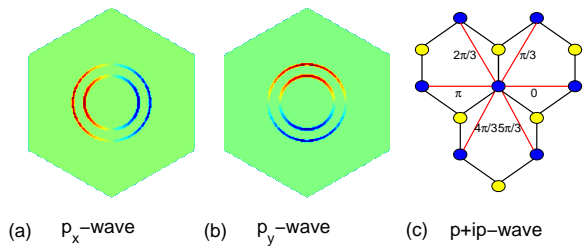}
\caption{(Color online) (a) and (b) show the $p_x$ and $p_y$ pairing symmetries in the $k$-space and (c) shows the phase of the $p+ip$ pairing symmetry on the honeycomb lattice in the real space.}
\label{Fig:pip}
\end{figure}

\begin{figure}[tbp]
\includegraphics[scale=0.45]{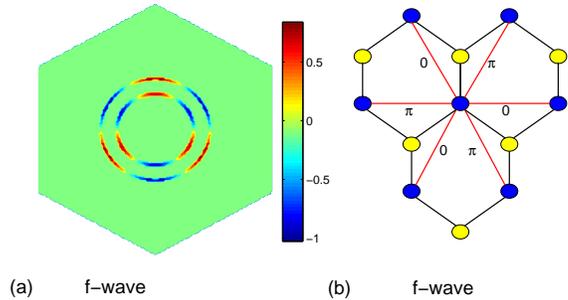}
\caption{
(a) shows the $f$ pairing in the $k$-space and (b) shows the phase of the $f$ pairing symmetries in the real space.}
\label{Fig:f}
\end{figure}

\begin{figure}[tbp]
\includegraphics[scale=0.45]{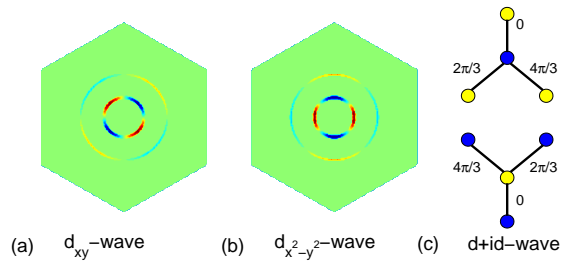}
\caption{(Color online) (a) and (b) show the $d_{xy}$ and $d_{x^{2}-y^{2}}$ pairing symmetries in the $k$-space and (c) shows the phase of the $d+id$ pairing symmetries in the real space.
}
\label{Fig:did}
\end{figure}

 Our results for $n$=0.1 and $n$=0.2 reveal that the leading pairing symmetries of the system at these low fillings are degenerate $p_x$ and $p_y$ doublets, as shown in Fig. \ref{Fig:pip}(a) and (b), which should be further mixed as $p_x\pm i p_y$ to minimize the ground state energy, as suggested by our further mean-field calculations on the effective Hamiltonian. Such a triplet pairing is mediated by the ferromagnetic-like spin fluctuations in the system, as shown in Fig. \ref{Fig:Spin}. The subleading pairing symmetries of the system at these low fillings are triplet $f$-wave shown in Fig. \ref{Fig:f}(a) for $n=0.1$ and singlet $d_{xy}$ and $d_{x^{2}-y^{2}}$ doublets (which should further be mixed as $d_{xy} \pm id_{x^{2}-y^{2}}$ to lower the energy) shown in Fig. \ref{Fig:did}(a) and (b) for $n=0.2$.

 Note that we have chosen such a small $U$ as $U=0.1t$ in our RPA calculations. For larger $U$ beyond its critical value $U_c$, the divergence of the spin susceptibility invalidate our RPA calculations for superconductivity. Physically, such a divergent spin susceptibility for $U>U_c$ may not necessarily lead to a magnetically-ordered state since the distribution of the susceptibility shown in Fig. \ref{Fig:Spin} does not possess a sharply peaked structure at particular momentum. Instead, the competition among different wave vectors may lead to paramagnetic behavior or short-ranged spin correlations which provide basis for the cooper pairing. We leave the study for the case of $U>U_c$ to the following DQMC approach, which is suitable for strong coupling problems.


\section {DQMC simulations}
The DQMC simulation is a powerful unbiased numerical tool to study the
physical properties of such strongly-correlated electronic systems as the Hubbard model.
The basic strategy of DQMC is to express the partition function as a high-dimensional integral over a set of
random auxiliary fields. The integral is then accomplished by
Monte Carlo techniques. For more technique details, we refer to
Refs.~\cite{Blankenbecler1981,Ma2010,MaReview2011}.

\begin{figure}[tbp]
\includegraphics[scale=0.18]{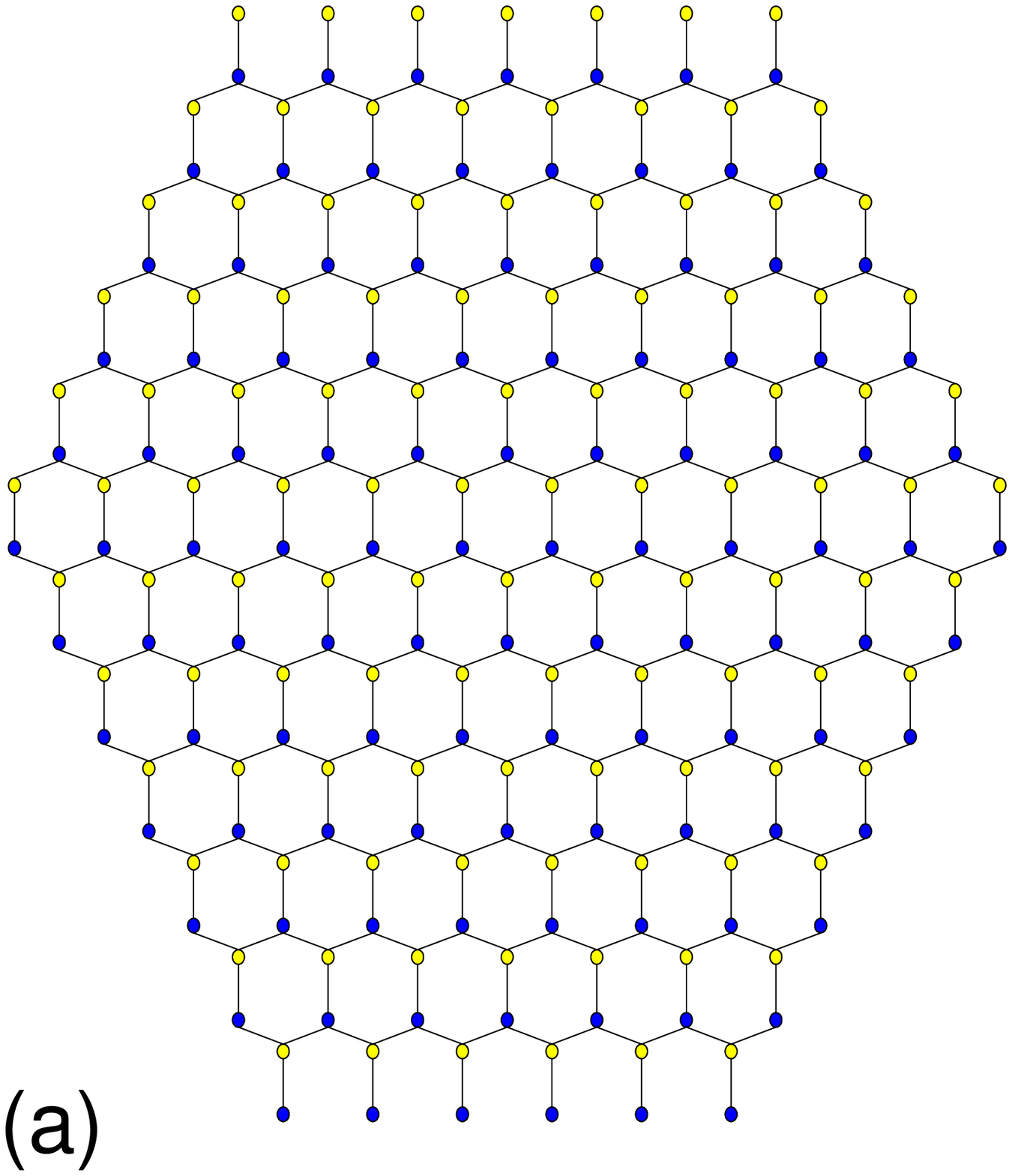}
\includegraphics[scale=0.18]{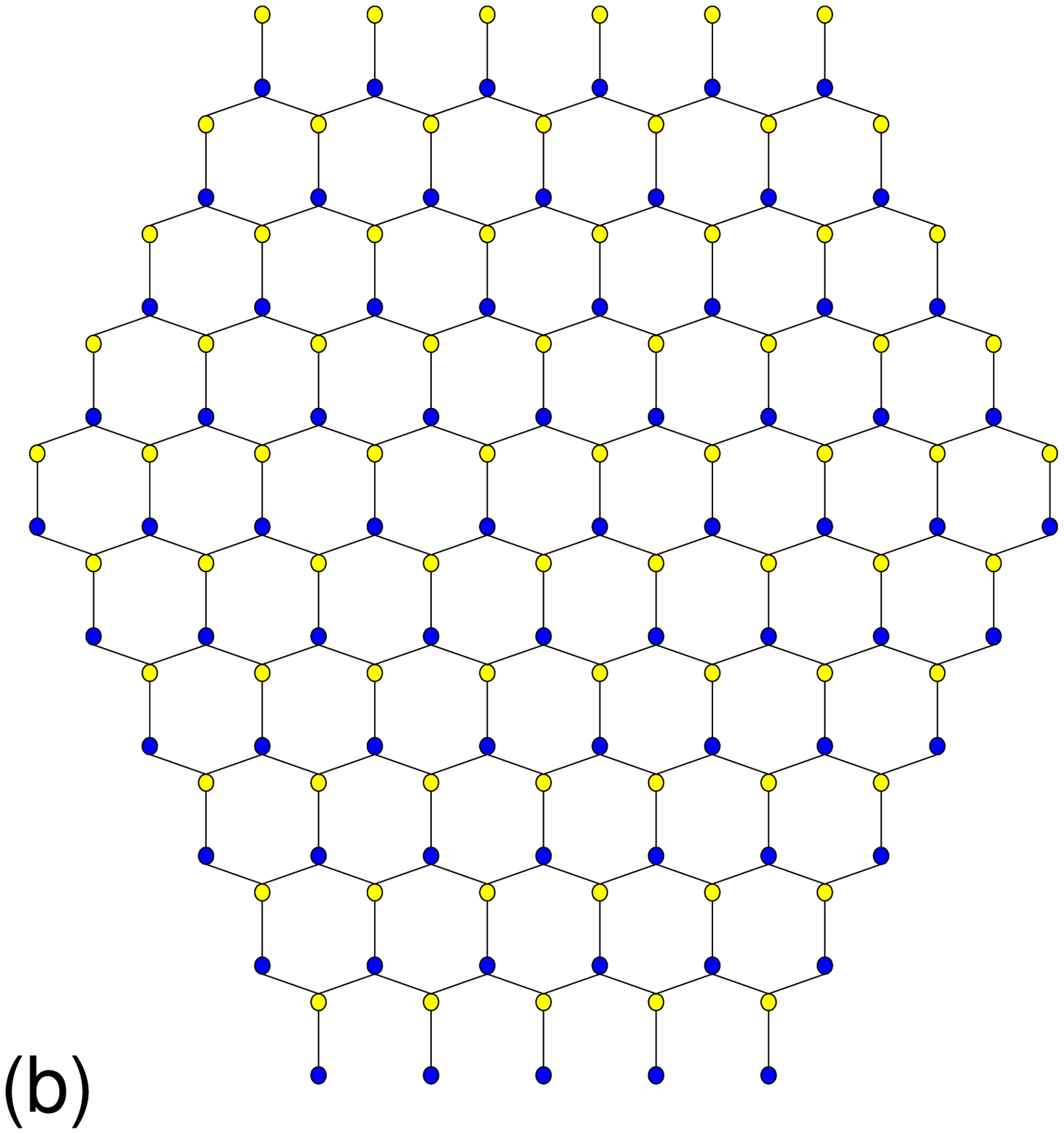}
\includegraphics[scale=0.18]{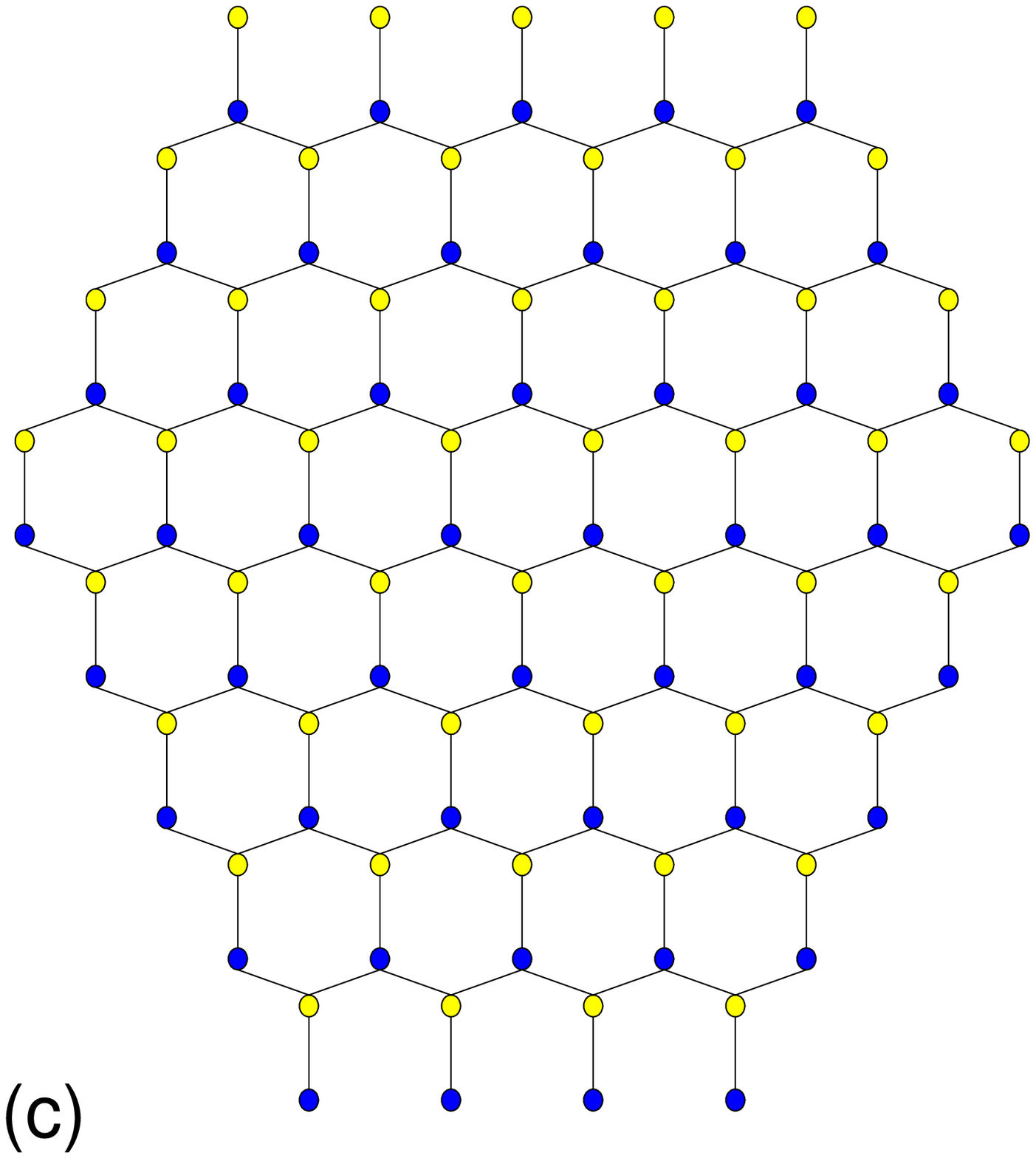}
\caption{The lattice geometries for the $2\times108$ (a), $2\times75$ (b) and $2\times48$ (c) honeycomb lattices.}
\label{Fig:lattice}
\end{figure}

To investigate the SC property, we compute the pairing
susceptibility,
\begin{equation}
P_{\alpha}\equiv\frac{1}{N_s}\sum_{i,j}\int_{0}^{\beta }d\tau \langle \Delta
_{\alpha }^{\dagger }(i,\tau)\Delta _{\alpha }^{\phantom{\dagger}%
}(j,0)\rangle.\label{sus}
\end{equation}
Here $\alpha$ stands for the pairing symmetry, and the corresponding pairing order parameter $\Delta_{\alpha }^{\dagger }(i)$\ is defined as
\begin{eqnarray}
\Delta_{\alpha }^{\dagger }(i)\ \equiv\sum_{l}f_{\alpha}^{*}
(\delta_{l})(c_{{i}\uparrow }c_{{i+\delta_{l}}\downarrow }\pm
c_{{i}\downarrow}c_{{i+\delta_{l}}\uparrow })^{\dagger},
\end{eqnarray}
where $f_{\alpha}(\delta_{l})$ is the form factor of the pairing
function, the vectors $\delta_{l}$ denote the bond connections, and ``$\pm$" labels triplet/singlet symmetries respectively. 

\begin{figure}[tbp]
\includegraphics[scale=0.425]{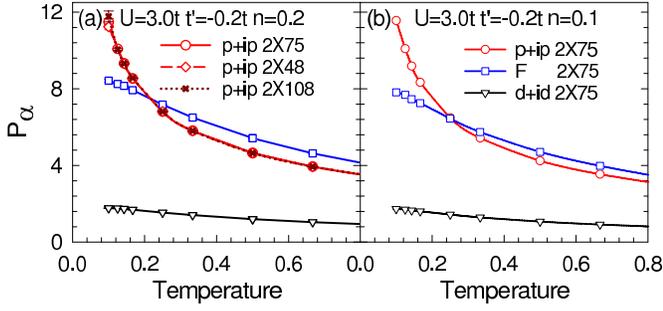}
\caption{(Color online) Pairing susceptibility $P_{\alpha}$ as a function
of temperature for different pairing symmetries with $U=3t$ at $n$=0.2 (a) and $n$=0.1 (b) on a $2 \times 75$ lattice (solid line). The $P_{p+ip}$ at $n$=0.2 on a $2 \times 48$ lattice (dash red line) and a $2 \times 108$ lattice are also shown (dotted red line) in (a). Here the units of temperature is $t$.}
\label{Fig:Pairing}
\end{figure}

Guided by the RPA results, three different pairing symmetries were investigated in the following DQMC studies, {\it i.e.} $p+ip$, $f$, and $d+id$ symmetries, whose form factors are illustrated in Fig. \ref{Fig:pip}(c), Fig. \ref{Fig:f}(b), and Fig. \ref{Fig:did}(b) respectively. These different pairing symmetries can be distinguished by their different phase shifts upon each 60$^\circ$ rotation, which are $\pi/3$, $2\pi/3$ and $\pi$ respectively. The NNN-bond $p+ip$ and $f$ wave triplet pairings shown possess the following form factors,
\begin{eqnarray}
\ f_{p+ip}(\delta_{l})=e^{i(l-1)
\frac{\pi }{3}},\ f_{f}(\delta_{l})=(-1)^{l},~l=1,\cdots,6,
\end{eqnarray}
and the NN-bond singlet $d+id$ pairing shown possesses the form factor
\begin{eqnarray}
\ f_{d+id}(\delta_{l})=e^{i(l-1)
\frac{2\pi }{3}},~l=1,2,3.
\end{eqnarray}
Note that the NN-bond pairing is prohibited in the $f$-symmetry. As for the $p+ip$ and $d+id$ ones, although pairings on both the NN-bond and the NNN-bond are allowed, our DQMC calculations show they are weaker (stronger) on the former than on the latter for the $p+ip$ ($d+id$) symmetry, reflecting the fact that the spin-fluctuations on the former are less ferromagnetic-like than those on the latter, consistent with our RPA calculations. We have also studied longer-range pairings by adding third and forth bond pairings in former factors, which turn out be much weaker 
than that of the NN-bond and NNN-bond presented above.

Our DQMC simulations of the system were performed at finite temperatures on a $2\times48$, a $2\times75$ and a $2\times108$ lattices with periodic boundary conditions. Here, each lattice we employed in simulations consists of two interpenetrating triangular sublattices with hexagonal shape such that it preserves most geometric symmetries of graphene, as shown in Fig.~\ref{Fig:lattice}. In each case, the total number of unit cells is $3L^2$ and the total number of lattice sites is $2\times3L^2$ with $L=$6, 5, or 4 in Fig.~\ref{Fig:lattice} (a), (b) and (c) respectively. Fig.~\ref{Fig:Pairing} shows the temperature dependence of the pairing susceptibilities for different pairing symmetries with electron filling $n$=0.2 (a) and $n$=0.1 (b) with $U=3t$.
Within the parameter range investigated, the pairing susceptibilities for various symmetries increase as the temperature is lowered, and most remarkably, the
$p+ip$ pairing symmetry dominates other ones at relatively low temperatures, consistent with the RPA results. In Fig.\ref{Fig:Pairing} (a), the pairing susceptibility $P_{p+ip}$ on a $2\times 48$ and a $2\times 108$ lattices are also shown, in comparison with that on the $2 \times 75$ lattice, from which one verifies negligible finite size effects.

\begin{figure}[tbp]
\includegraphics[scale=0.375]{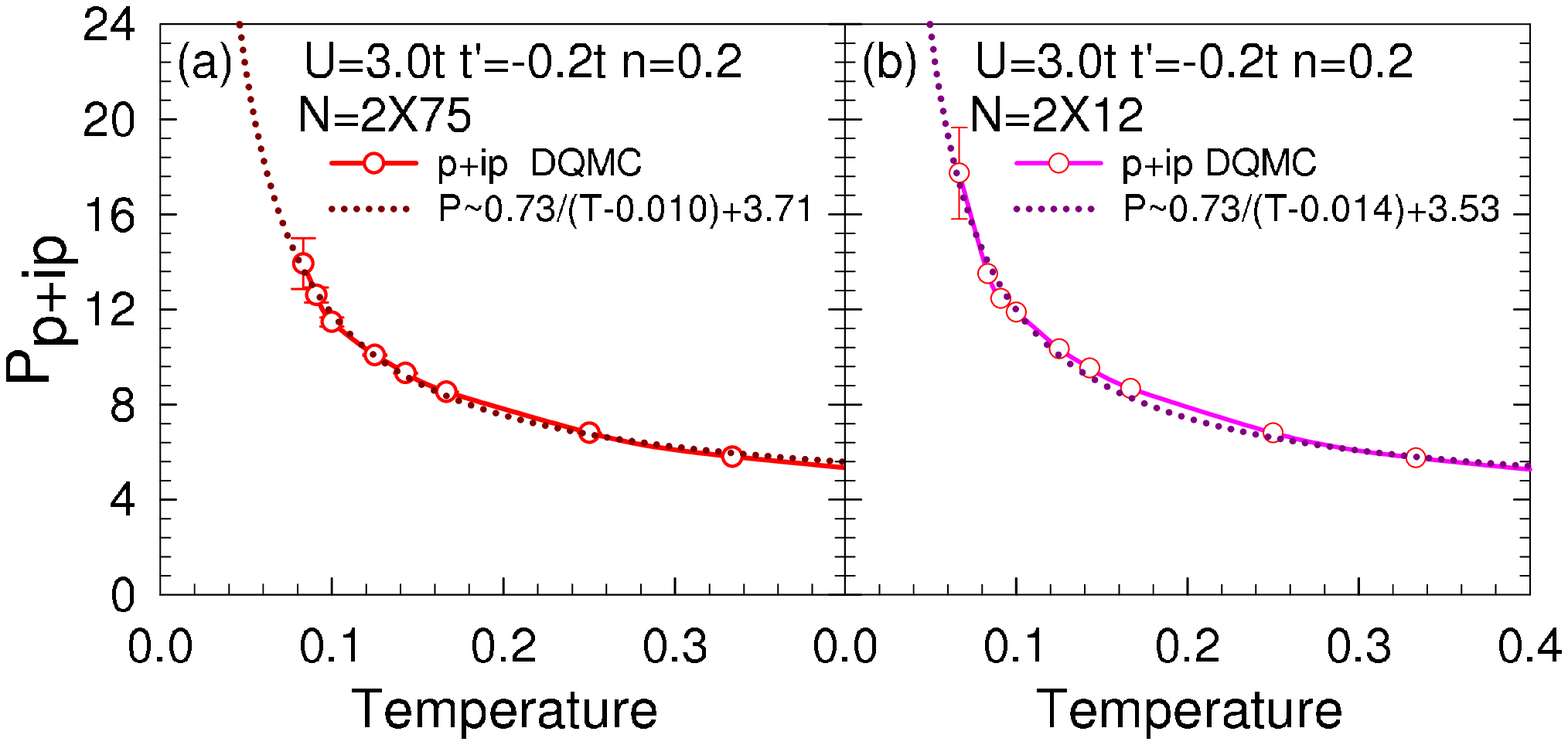}
\caption{(Color online) Pairing susceptibility $P_{p+ip}$ as a function
of temperature with $U=3t$ and $n$=0.2 for a $2 \times 75$ lattice (a) and a $2 \times 12$ lattice (b) (solid line). The fitting data are also shown as dashed lines.}
\label{Fig:Tc}
\end{figure}

The superconducting transition occurs as the pairing susceptibility diverges. However, DQMC simulations encounter the notorious minus problem in this doped system as well; consequently the lower the temperature used in DQMC, the larger the error bar is. In Fig. \ref{Fig:Tc}, we have simulated the system to the lowest temperature at our best while keep a reasonable error bar.  The lowest temperature for the $2 \times 75$ lattice is $t/12$ and the lowest temperature for the $2 \times 12$ lattice is $t/15$. Within our numerical results, We fit the DQMC data with a formula of $P=a/(T-T_c)+b$, as shown (dashed lines) in Fig. \ref{Fig:Tc} and then we extrapolate to obtain the $T_c$.
The fitting agrees with the DQMC data reasonably well. From this fitting, one may estimate a $T_c$ of about $\sim 0.01t$, which is roughly $\sim 200$K.

\begin{figure}[tbp]
\includegraphics[scale=0.375]{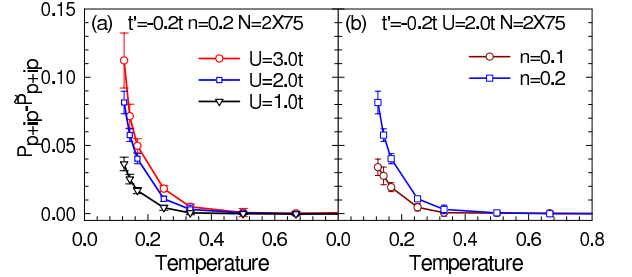}
\caption{(Color online) The intrinsic pairing interaction
$P_{p+ip}-\widetilde{P}_{p+ip}$ as a function of temperature for different $U$ (a) and  different $n$ (b) on a $2 \times 75$ lattice.}
\label{Fig:FigU}
\end{figure}

In order to extract the intrinsic pairing interaction in our finite system, one should subtract from $P
_{\alpha}$ its uncorrelated single-particle contribution $\widetilde{P}
_{\alpha}$, which is achieved by replacing $\langle
c_{{i}\downarrow }^{\dag }c_{{j}\downarrow }c_{i+\delta_{l}\uparrow}^{\dag}
c_{j+\delta_{l'}\uparrow}\rangle $ in Eq. (\ref{sus}) with $\langle c_{{i}\downarrow }^{\dag
}c_{{j}\downarrow }\rangle \langle c_{i+\delta_{l}\uparrow }^{\dag }
c_{j+\delta_{l'}\uparrow }\rangle $. Clearly in Fig. \ref{Fig:FigU}, the intrinsic pairing interaction $P_{p+ip}-\widetilde{P}_{p+ip}$ shows qualitatively the same temperature dependence as that of $P_{p+ip}$, which is positive and increases with the lowering of temperature. Such a temperature dependence of $P_{\alpha}-\widetilde{P}_{\alpha}$ suggests effective attractions generated between electrons and the instability toward SC in the system at low temperatures. Moreover, Fig. \ref{Fig:FigU}(a) shows that the intrinsic pairing interaction for $p+ip$ symmetry enhances with larger $U$, indicating the enhanced pairing strength with the enhancement of the electron correlations. As for the other two pairing symmetries shown, our DQMC results yield negative intrinsic pairing interactions, reflecting the fact that the realization of the $p+ip$ symmetry at low temperatures will suppress other competing pairing channels.


\section {Conclusions and discussions}
 We have performed combined RPA analysis and DQMC calculations for the low-filled honeycomb Hubbard model with weak and strong repulsive U respectively. Both studies show that the triplet $p+ip$ SC occurs as the ground state of our model system of low-filled graphene. 
Besides graphene, the results obtained here also apply to other isostructure materials, such as silicene\cite{silicene} and germanene\cite{germanene}. Furthermore, by trapping some fermionic cold atoms into an optical lattice, one may also be able to simulate the Hubbard-model on a honeycomb lattice studied here\cite{Zhu2007,Wu2007,coldatom}, with tunable parameters and dopings, which is expected to realize the triplet $p+ip$ superfluidity.

{\bf Acknowledgement}: We would like to thank Zhong-Bin Huang, Yuigui Yao, Yu-Zhong Zhang, and Su-Peng Kou for stimulating discussions. This work is supported in part by NSFC (Grant Nos. 11104014, 11274041, 11374034, and 11334012), by Research Fund for the Doctoral Program of Higher Education of China 20110003120007 and SRF for ROCS (SEM) (T.M.), by the NCET program under the grant No. NCET-12-0038 (F.Y.), and by the Thousand-Yound-Talent Program of China (H.Y.).


\begin{thebibliography}{35}%










\bibitem{graphene}K.S. Novoselov, A. K. Geim, S. V. Morozov, D. Jiang,
Y. Zhang, S. V. Dubonos, I. V. Gregorieva, and A. A.
Firsov, Science {\bf 306}, 666 (2004).

\bibitem{dirac} For a review, see A. H. Castro Neto, F. Guinea, N. M. R.
Peres, K. S. Novoselov and A. K. Geim, Rev. Mod. Phys.
{\bf 81}, 109 (2009).

\bibitem{raghu10} S. Raghu, S. A. Kivelson, and D. J. Scalapino, Phys. Rev. B {\bf 81}, 224505 (2010).
\bibitem{Nandkishore} R. Nandkishore, R. Thomale and A. V. Chubukov, arXiv:1401.5485.

\bibitem{chubukov} R. Nandkishore, L. S. Levitov, and A. V. Chubukov, Nat.
Phys. {\bf 8}, 158 (2012).

\bibitem{wang} W.-S. Wang, Y.-Y. Xiang, Q.-H. Wang, F. Wang, F. Yang, and D.-H. Lee, Phys. Rev. B {\bf 85}, 035414 (2012).

\bibitem{thomale} M. L. Kiesel, C. Platt, W. Hanke, D. A. Abanin, and R. Thomale, Phys. Rev. B {\bf 86}, 020507(R) (2012).

\bibitem{did1} A. M. Black-Schaffer, S. Doniach, Phys. Rev. B {\bf 75,} 134512 (2007).

\bibitem{did2}J. Gonz\'{a}lez, Phys. Rev. B {\bf 78,} 205431 (2008).

\bibitem{did3} S. Pathak, V. B. Shenoy, G. Baskaran, Phys. Rev. B {\bf 81,} 085431 (2010).

\bibitem{Platt} C. Platt, W. Hanke, and R. Thomale, arXiv:1310.6191.

\bibitem{Qi-Zhang} X.-L. Qi and S.-C. Zhang, Rev. Mod. Phys. {\bf 83}, 1057 (2011).

\bibitem{Hasan-Kane} M. Z. Hasan and C. L. Kane, Rev. Mod. Phys. {\bf 82}, 3045 (2010).

\bibitem{litao11} T. Li, arXiv:1103.2420 (2011).

\bibitem{martin08} I. Martin and C. D. Batista,  Phys. Rev. Lett. {\bf 101}, 156402 (2008).

\bibitem{yao13} H. Yao and F. Yang, arXiv:1312.0077.

\bibitem{chenxi} X. Chen, Y. Yao, H. Yao, F. Yang and J. Ni, arXiv:1404.3346.

\bibitem{FP} J. Jung and A. H. MacDonald, Phys. Rev. B {\bf 87}, 195450 (2013).
\bibitem{exp1} A. Kretinin, G. L. Yu, R. Jalil, Y. Cao, F. Withers, A. Mishchenko, M. I. Katsnelson, K. S. Novoselov,
A. K. Geim, and F. Guinea, Phys. Rev. B {\bf 88}, 165427 (2013)
\bibitem{exp2} R. S. Deacon, K. C. Chuang, R. J. Nicholas, K. S. Novoselov, and A. K. Geim, Phys. Rev. B {\bf 76}, 081406 (2007).
\bibitem{doping_high} J. T. Ye, S. Inoue, K. Kobayashi, Y. Kasahara, H. T. Yuan, H. Shimotani and Y. Y. Iwasa, Nature
Mater. 9, 125-128 (2010).
\bibitem{Ma2010} T. Ma, F. M. Hu, Z. B. Huang, and H. Q. Lin, Appl. Phys. Lett. {\bf 97}, 112504 (2010).

%

\bibitem{Scalapino89} N. E. Bickers, D. J. Scalapino, and S. R. White, Phys. Rev. Lett. {\bf 62}, 961 (1989).

\bibitem{Scalapino-RMP} D. J. Scalapino, Rev. Mod. Phys. {\bf 84}, 1383 (2012).

\bibitem{RPA1} T. Takimoto, T. Hotta, and K. Ueda, Phys. Rev. B {\bf 69}, 104504 (2004); K. Yada and H. Kontani, J. Phys. Soc. Jpn. {\bf 74}, 2161 (2005).

\bibitem{Kubo2007} K. Kubo, Phys. Rev. B {\bf 75}, 224509 (2007); K. Kuroki, S. Onari, R. Arita, H. Usui, Y. Tanaka, H. Kontani, and H. Aoki, Phys. Rev. Lett. {\bf 101}, 087004 (2008).

\bibitem{Graser2009} S. Graser, T. A. Maier, P. J. Hirschfeld, and D. J. Scalapino, New Journal of Physics {\bf 11}, 025016 (2009).

\bibitem{RPA2} F. Liu, C.-C. Liu, K. Wu, F. Yang, and Y. Yao, Phys. Rev. Lett. {\bf 111}, 066804 (2013); L.-D. Zhang, F. Yang and Y. Yao, arXiv:1309.7347

\bibitem{Blankenbecler1981} R. Blankenbecler, D. J. Scalapino, and R. L. Sugar, Phys. Rev. D {\bf 24}, 2278 (1981).

\bibitem{MaReview2011} T. Ma, F. M. Hu, Z. B. Huang, and H.-Q. Lin,  $Horizons$ $in$ $World$ $Physics$, {\bf 276}, Chapter 8, Nova Science Publishers, Inc, 2011.

\bibitem{silicene} B. Lalmi, H. Oughaddou, H. Enriquez, A. Karae, S. Vizzini, B. Ealet, and B. Aufray, Appl. Phys. Lett. \textbf{97}, 223109 (2010).
\bibitem{germanene} M.E. D¨¢vila, L. Xian, S. Cahangirov, A. Rubio, G. Le Lay, arXiv:1406.2488.
\bibitem{Zhu2007} S.-L. Zhu, B.G. Wang, and L.-M. Duan, Phys. Rev. Lett. {\bf 98}, 260402 (2007).

\bibitem{Wu2007} C. J. Wu, D. Bergman, L. Balents, and S. Das Sarma, Phys. Rev. Lett. {\bf 99}, 070401 (2007).

\bibitem{coldatom} L. Tarruell, D. Greif, T. Uehlinger, G. Jotzu, and T.
Esslinger, Nature {\bf 483}, 302 (2012).

\end{thebibliography}


\end{document}